\documentclass[sigconf]{acmart}

\usepackage{siunitx}

\usepackage{textcomp}
\usepackage{tabularx}

\usepackage{booktabs, siunitx, threeparttable}
\usepackage{adjustbox}
\usepackage{balance}

\usepackage{float}

\usepackage{threeparttable}
\usepackage[utf8]{inputenc}
\usepackage{textgreek}
\AtBeginDocument{%
  }

\setcopyright{acmlicensed}
\copyrightyear{2025} 
\acmYear{2025} 
\setcopyright{acmlicensed}\acmConference[MM '25]{Proceedings of the 33rd ACM International Conference on Multimedia}{October 27--31, 2025}{Dublin, Ireland}
\acmBooktitle{Proceedings of the 33rd ACM International Conference on Multimedia (MM '25), October 27--31, 2025, Dublin, Ireland}
\acmDOI{10.1145/3746027.3754580}
\acmISBN{979-8-4007-2035-2/2025/10}

\begin{document}

\sisetup{round-mode=places, round-precision=3, detect-weight=true, detect-inline-weight=math}

\title{Generative Multi-Sensory Meditation: Exploring Immersive Depth and Activation in Virtual Reality}

\author{Yuyang Jiang}
\affiliation{
  \institution{The Hong Kong University of Science and Technology}
  \city{Hong Kong}
  \country{China}}
\affiliation{%
 \institution{The Hong Kong University of Science and Technology, Guangzhou}
 \city{Guangdong}
  \country{China}}
\email{yjiang257@connect.hkust-gz.edu.cn}

\author{Binzhu Xie}
\affiliation{%
  \institution{The Chinese University of Hong Kong}
  \city{Hong Kong}
  \country{China}}
\email{bzxie@cse.cuhk.edu.hk}

\author{Lina Xu}
\affiliation{%
 \institution{The Hong Kong University of Science and Technology, Guangzhou}
 \city{Guangdong}
  \country{China}}
\email{lxu582@connect.hkust-gz.edu.cn}

\author{Xiaokang Lei}
\affiliation{%
  \institution{The Hong Kong University of Science and Technology, Guangzhou}
  \city{Guangdong}
  \country{China}}
\email{xlei688@connect.hkust-gz.edu.cn}

\author{Shi Qiu}
\affiliation{%
  \institution{The Chinese University of Hong Kong}
  \city{Hong Kong}
  \country{China}
  }
\email{shiqiu@cse.cuhk.edu.hk}

\author{Luwen Yu}
\affiliation{%
  \institution{The Hong Kong University of Science and Technology, Guangzhou}
  \city{Guangdong}
  \country{China}}
\email{luwenyu@hkust-gz.edu.cn}

\author{Pan Hui}
\authornote{Corresponding author.}
\affiliation{%
  \institution{The Hong Kong University of Science and Technology}
  \city{Hong Kong}
  \country{China}}
\affiliation{%
 \institution{The Hong Kong University of Science and Technology, Guangzhou}
 \city{Guangdong}
  \country{China}}
\email{panhui@ust.hk}
\renewcommand{\shortauthors}{Yuyang Jiang et al.}


\begin{abstract}
Mindfulness meditation has seen increasing applications in diverse domains as an effective practice to improve mental health. However, the standardized frameworks adopted by most applications often fail to cater to users with various psychological states and health conditions. This limitation arises primarily from the lack of personalization and adaptive content design. To address this, we propose MindfulVerse, an AI-Generated Content (AIGC)-driven application to create personalized and immersive mindfulness experiences. By developing a novel agent, the system can dynamically adjust the meditation content based on the ideas of individual users. Furthermore, we conducted exploratory user studies and comparative evaluations to assess the application scenarios and performance of our novel generative meditation tool in VR environments. The results of this user study indicate that generative meditation improves neural activation in self-regulation and shows a positive impact on emotional regulation and participation. Our approach offers a generative meditation procedure that provides users with an application that better suits their preferences and states.
\end{abstract}




\maketitle

\section{Introduction}
Mindfulness is the practice of focused, nonjudgmental awareness of the present moment~\cite{liu2022effectiveness,matis2020mindfulness}. Initially popularized in clinical settings through programs such as Mindfulness-Based Stress Reduction (MBSR)~\cite{Alsubaie2017}, mindfulness shares similarities with meditation but emphasizes practical attention training. Research shows its benefits for mental health, including reduced anxiety and improved emotional regulation. Recently, virtual reality (VR) has expanded mindfulness-related applications~\cite{Ma2025,riva2005}, offering immersive mindfulness training for stress relief, workplace wellness, and therapeutic interventions. For example, medical institutions use immersive simulated environments to assist patients in relaxation training~\cite{Huberty2021A,Shen2020}. Corporations and institutions implement VR-based programs as part of employee wellness initiatives, providing stress management solutions~\cite{khoury2013mindfulness}.

Although mindfulness and meditation media often claim initial positive effects, their impact tends to diminish over time~\cite{Nina2021,fincham2023effects,lazar2013neurobiology}. This decline is largely due to psychological habituation: Users become accustomed to repetitive structured content, leading to reduced engagement and weaker therapeutic outcomes. AI-Generated Content (AIGC) technology presents both opportunities~\cite{10896112} and challenges~\cite{Silva2024Could,williams2022supporting}. AIGC-based techniques such as AI-generated music~\cite{Yu2023Developments,lin2024harmony}, visuals~\cite{Ameta2023,zhu2023genimage}, and responsive dialogues~\cite{Kuhail2024Human} offer new ways to enhance personalization and sustain engagement.

Despite these advancements, the psychological effects of immersive generative meditation environments remain underexplored. 
In particular, while previous research has focused on static VR meditation environments~\cite{Xu2025,kumar2024vrzm}, the potential of dynamically generated user-specific environments to improve meditation effectiveness has not been thoroughly investigated~\cite{moseley2017deep}.
Furthermore, traditional methods of assessing cognitive and psychological outcomes, which rely primarily on behavioral data or self-reports, do not capture the complexity of the brain response to such experiences~\cite{Shapiro1992}. Based on extensive user research and identified demands, we introduce ``MindfulVerse" -- a prototype that uses generative AI for multimodal content generation, offering immersive and personalized mindfulness meditation experiences with AI Agent.

To evaluate the effectiveness of AI-generated personalized meditation in immersive VR, our study focuses on three key questions. First, our objective is to investigate how AI-generated and personalized meditation content differentially activate meditation-related brain regions (RQ1). Second, we explore how four types of multimodal personalized meditation (image-based, music-based, guided-word-based, and integrated) differentially influence emotional responses (RQ2). In addition, we examine whether generative meditation experiences lead to significant differences in perceived immersion, engagement, and overall user satisfaction compared to static content, combining qualitative and quantitative measures to assess user experience in VR (RQ3).

Oriented by research questions, the experimental design is structured to explore the experiential impacts of AI-generated personalized meditation in immersive VR. Our experiment adopts a mixed research method with user study and employs functional near-infrared spectroscopy (fNIRS) to observe brain activation.
To evaluate the effectiveness of our approach in improving mindfulness reflection and emotional regulation, we conducted a user study with 30 participants. Each participant is asked to rate each generation mode and provide justification for whether personalized integrated meditation meets their individual needs.

 Our contributions are as follows: (1) MindfulVerse is designed for generative meditation, located at the intersection of mindfulness practice and adaptive virtual experience design. (2) Our study compares multimodal generative mindfulness meditation with single-modal meditation methods and explores their unique differences in the brain activity of users. (3) Through systematic user evaluation, this work highlights the value of multimodal coherence in supporting presence perception and emotional resonance, providing practical information for the design of VR meditation systems.

\vspace{-10pt}

\section{Related Work}
\textbf{\textit{  Mindfulness and Meditation in Mental Health.  }} Mindfulness practice has been shown to support emotional regulation, improve mental health, control of attention, and improve the neurophysiological stress response in a variety of populations~\cite{Bellehsen2021A,tang2015neuroscience,boccia2015meditative}. These benefits contribute to its growing adoption in therapeutic, educational, and wellness contexts. Existing mindfulness systems typically rely on structured programs such as Mindfulness-Based Cognitive Therapy (MBCT)~\cite{Day2017}, often delivered through pre-recorded audio guidance~\cite{Goldberg2019,tan2025} or instructor-led sessions. Although effective in the short term, prolonged exposure to repetitive guided content often triggers sensory adaptation, reducing neural responsiveness in regions associated with emotion and attention~\cite{Yang2025Comparative}. Although recent findings confirm these effects, current delivery methods rarely adjust to individual adaptation patterns~\cite{Shapiro1992,Stecher2021Using}, and few offer mechanisms to modulate content to maintain neural and emotional engagement over extended periods. Meditation activates eight key brain regions~\cite{Spijkerman2016}, with specific areas varying according to type of practice~\cite{Bergen2021}. This indicates that effective meditation can help improve cognitive control and attention maintenance~\cite{Kwak2020Enhanced}. However, prolonged engagement with identical mindfulness routines results in reduced activation across key brain regions~\cite{Izzetoglu2020,Young2018}. Functional imaging reveals that this neural attenuation extends beyond isolated areas to larger network-level reductions in connectivity and responsiveness~\cite{Garrison2015}. These patterns suggest the brain becomes accustomed to repetitive stimuli, undermining the intensity and effectiveness of practice even when behavioral adherence remains stable. Meditation experiences can stagnate or regress without personalized strategies to sustain neural engagement.

\textbf{\textit{  VR and Immersion.  }} In particular, VR enables embodied experiences that enhance presence and reduce external distraction~\cite{Chandrasiri2019,navarro2017meditation,navarro2019evaluation}. It offers dynamic environments that respond to user input or physiological data in real time~\cite{2024Exploring,Wang2023,Pyjas2022Storytelling}. In this context, Feinberg et al. designed ZenVR, comprising eight VR meditation sessions to help beginners develop meditation skills and mindfulness~\cite{Feinberg2022}. Ameta et al. tested a 360° VR-based mandala meditation environment and found significant improvements in stress, mood, and cognition~\cite{aggarwal2023artificial}. Kaplan-Rakowski et al. showed that VR-based meditation leads to greater anxiety reduction than traditional video formats~\cite{verma2023attentionet,norouzi2024precision}. This opens new design possibilities for personalized meditation beyond pre-recorded scripts~\cite{Peng2023}. However, most current applications remain static and lack evaluation on sustained attention, emotional regulation, or neural engagement.

\begin{figure*}[hbpt]
    \centering
    \includegraphics[width=\linewidth]{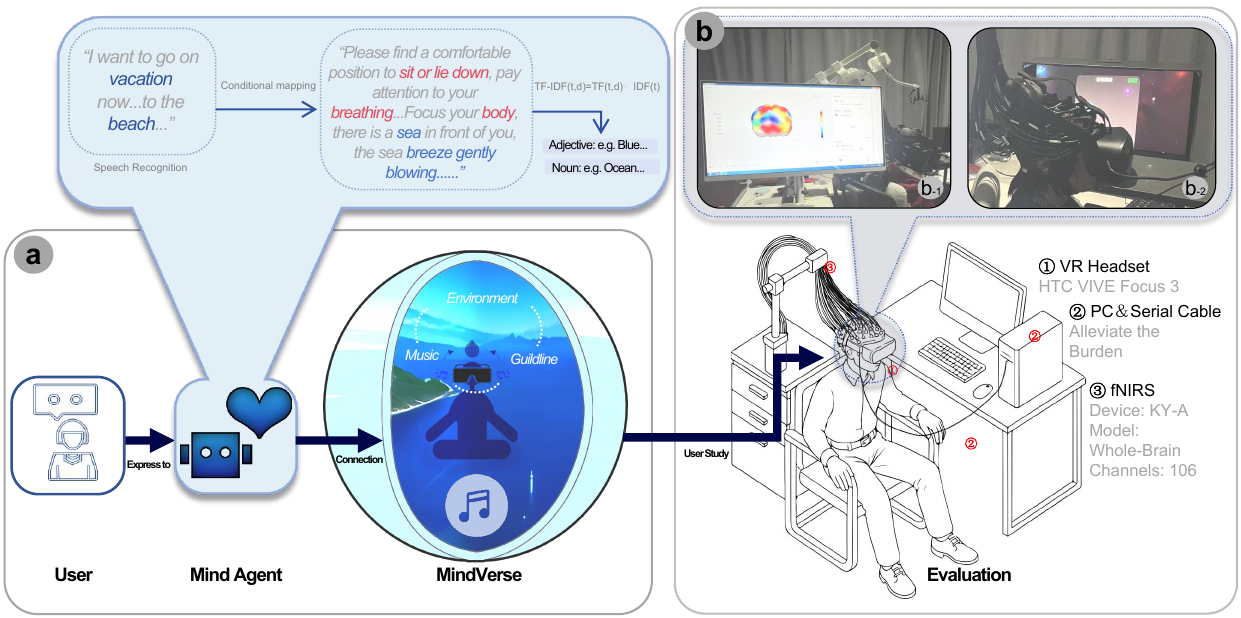}
    \vspace{-20pt}
    \caption{(a) The \textit{\textbf{MindfulVerse}} interaction pipeline. Users verbally express their mental state or intention, which drives the generation of personalized meditation content across three dimensions: environment, music, and guided prompts, forming an adaptive VR mindfulness experience. (b) Experimental protocol using fNIRS. Participants engage with the generated meditation in immersive VR while whole-brain hemodynamic activity is recorded.}
    \label{fig:tech}
    \vspace{-10pt}
\end{figure*}

\textbf{\textit{  AI-Generated Personalized Content.  }}
Digital mindfulness platforms are increasingly experimenting with generative content, adaptive interfaces, and immersive formats. Applications that incorporate mindfulness, relationship building, and self-insight training~\cite{Stecher2021Using}, and have been shown to effectively reduce psychological distress~\cite{Wang2023,wang2022reducing,fu2023loop} and enhance social connections~\cite{Goldberg2020,goldberg2022empirical}. These studies provide guided meditation audio, timers, recordings, scenario-based exercises, and mindfulness reminders. With technological advancement, the gradual integration of AI and physiology has enhanced user experience~\cite{Chavez2020,mistry2020meditating} and practice effectiveness~\cite{Simo2021}. However, some studies have also pointed out that many applications are driving the increasing instrumentalization of AI applications while overlooking its deeper impact on consciousness and the nervous system. Most existing studies rely solely on self-report scales to assess the effectiveness of meditation, lacking objective neuroscience evidence --- the personalized adaptation effect of the generated content remains unvalidated.

\section{Generative Mindfulness Tools in VR}
\subsection{Survey on Requirements}
In order to obtain better guidance for meditation and the requirements for prototype design, we recruited respondents (N=61) for the questionnaire through the platform they use for meditation, and most of the participants reported a relatively frequent practice. The questionnaire collected demographic information, meditation user rating, and an open survey of existing meditation tools (in the Appendix document). The design requirements and insights for generative meditation tools are derived from the following user needs:

\textbullet\ \textbf{Seek fresh and varied content} - inspires the development of systems capable of personalized content.

\textbullet\ \textbf{Encounter uncertainties during the experience} – highlights the importance of designing interactive agents and support to enhance user guidance in the meditation process.

\textbullet\ \textbf{Diverse emotional states and shifting} - calls for meditation systems that offer multiple styles and adaptable interaction modes to support user emotional regulation.



\subsection{Mindfulness Agent Design}
We designed the mindfulness agent to provide a personalized meditation experience by converting structured mindfulness content into high-quality guidance words~\cite{best2022freely,mitsea2023virtual}. These guidance words are generated under the constraints of a mindfulness paradigm, ensuring that the content can effectively guide users while avoiding the generation of inappropriate or irrelevant language.

\textbf{\textit{ Training Preparation.  }} The training data set is derived from real mindfulness classes, encompassing various meditation techniques, relaxation exercises, and cognitive strategies, similar to the paradigms and media content used in established mindfulness-based interventions~\cite{kabat2009full}. 

\textbf{\textit{Prompt Generation.}}  The core of effective meditation guidance lies in the quality of the guidance words. To achieve this, a specialized prompt generator is trained to convert user-provided keywords into initial prompts~\cite{Xue2024Mult}. This module is designed to produce coherent and contextually relevant prompts that serve as the foundational input for subsequent processing. The generated prompts are further refined through a quality assurance model, which enhances their quality by ensuring coherence, clarity, and contextual relevance.

\textbf{\textit{Interaction.}} To ensure consistency and precision, a QA refinement module is employed to improve the quality of the generated prompts. By refining initial prompts through a fine-tuned QA model~\cite{khashabi2020unifiedqa}, the system produces guidance words that are clear, contextually relevant, and suitable for generating multimedia content.

\textbf{\textit{Technique Details.}} We fine-tuned Qianfan (based on ERNIE-Speed-8K) on the publicly available Meditation-miniset v0.2 corpus (20 k curated QA pairs). During the fine-tuning process, every training example was first processed through Qwen-2.5~\cite{qwen2025qwen25technicalreport} to simulate our AIGC environment (``argument structure''), and all resulting outputs were manually reviewed by annotators (N = 8, 24-30 years, psychological professionals) to ensure that they exhibit the desired AIGC characteristics while excluding any sensitive content. Here we post the machine and the latency: Prompt-Generation Latency: Music 580 ms; Speech 430 ms; Visual 720 ms; Backend GPU: NVIDIA A100 on Baidu Cloud and related content.

\subsection{Mindfulness Tool Pipeline}
\label{sec3.2}
The interaction pipeline of the generative mindfulness tool is centered on a core concept: the use of a \textit{guiding keyword}, aligning with principles found in traditional mindfulness practices. As shown in Figure~\ref{fig:tech}, these generation modes correspond to four types of mindful guidance theory: \textbf{Guided Imagery Mindfulness~\cite{Bergen2021}}, \textbf{Sensory Mindfulness~\cite{finck2023multisensory}}, \textbf{Environmental Mindfulness~\cite{meaden2024environmental}}, and \textbf{Integrated Mindfulness~\cite{proulx2003integrating}}.

Interaction begins when a user articulates a desired meditative state or intention through voice input. This is transcribed using Tencent Automatic Speech Recognition (ASR)~\footnote{https://www.tencentcloud.com/products/asr} and mapped to a structured guide keyword. Based on this keyword, an initial meditation script is generated by the system's core agent. To improve clarity and contextual fit, this script is then refined through a QA-enhanced feedback loop using the Qianfan module~\footnote{https://github.com/baidubce/bce-qianfan-sdk}.

Once the guidance content is finalized, it triggers a cascade of multimedia generation processes: structured verbal instructions (via Qianfan), ambient soundscapes (via Suno~\footnote{https://www.suno.com}), and immersive 360° visual scenes (via Skybox~\footnote{https://skybox.blockadelabs.com/}). Each element is coherently generated in alignment with the guiding keyword to maintain narrative and sensory consistency. Unity's cross-platform framework enables flexible deployment across desktop, mobile, and VR environments.

\section{User Study and Evaluation}
To comprehensively explore and assess the neuro-cognitive, emotional, and experiential impacts of generative mindfulness in VR, we designed a multimodal framework integrating neurophysiological measures and self-reports.



\subsection{Preparation}
A cohort of participants was recruited through campus-wide advertisements, following approval of the Institutional Review Board (IRB No.23 in HSP-2024-0054). Inclusion criteria required documented prior experience with VR systems, and individuals with a history of motion sickness, neurological conditions, or contraindications to prolonged use of head mounted displays were excluded (see Supplemental Materials for complete criteria).

\textbf{\textit{ Participants.}} All participants provided their written informed consent and were reminded of their right to withdraw at any time. Those who completed the full study received a participation voucher. During the experiment, four participants (IDs: M06, F05, M16, F08) withdrew before completing all sessions, resulting in a final sample of \textbf{30 participants} (including 16 males and 14 females). The final cohort ranged in age from 18 to 38 years (\(M=25.93\), \(SD=5.92\)). All participants were native Chinese with no reported psychiatric or neurological disorders.

\textbf{\textit{Laboratory Environment and Settings.}} All sessions were carried out in a quiet, controlled laboratory environment to minimize external interference (see Figure~\ref{fig:tech}). The experimental setup comprised two synchronized systems: a connected VR platform (HTC VIVE Focus 3, HTC) and a whole-brain functional near-infrared spectroscopy (fNIRS) system (KY-A, Wuhan YIRUIDE Group) with 106 channels. Throughout the experiment, the fNIRS system continuously recorded cortical hemodynamic responses, allowing the subsequent analysis of task-related brain activity in different meditation modalities.

 \vspace{-5pt}
\begin{figure}[H]
    \centering
    \includegraphics[width=1\linewidth]{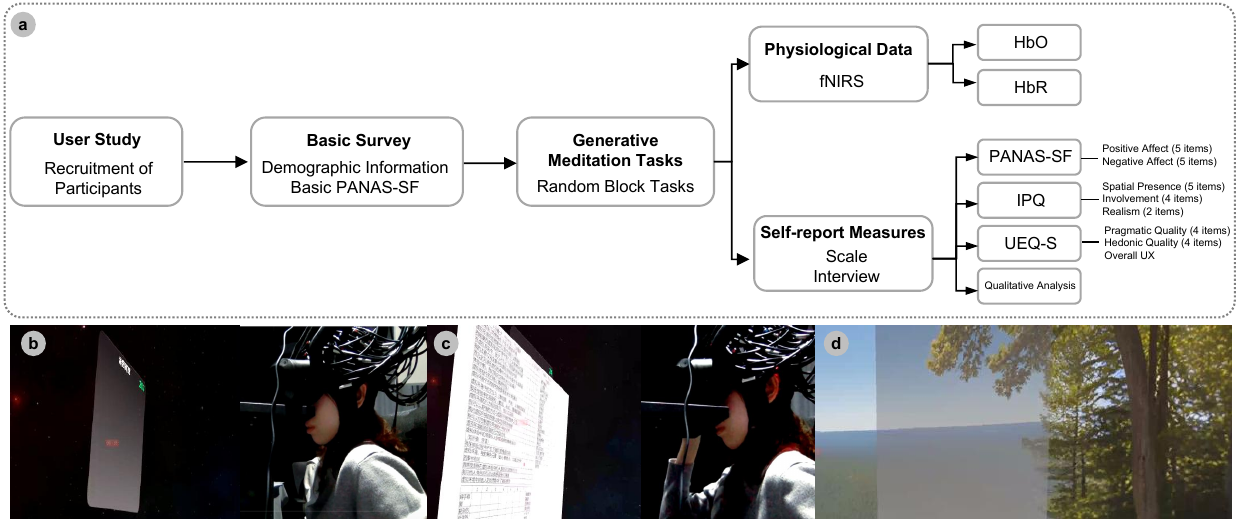}
    \vspace{-10pt}
    \caption{Procedure of the study. (a) User‑study workflow. (b) Initial setup/entry state. (c) In‑system questionnaire completion. (d) Task buffer/transition scene.}
    \label{sacle}
        \vspace{-15pt}
\end{figure}

\begin{figure*}[h]
    \centering
    \includegraphics[width=1\linewidth]{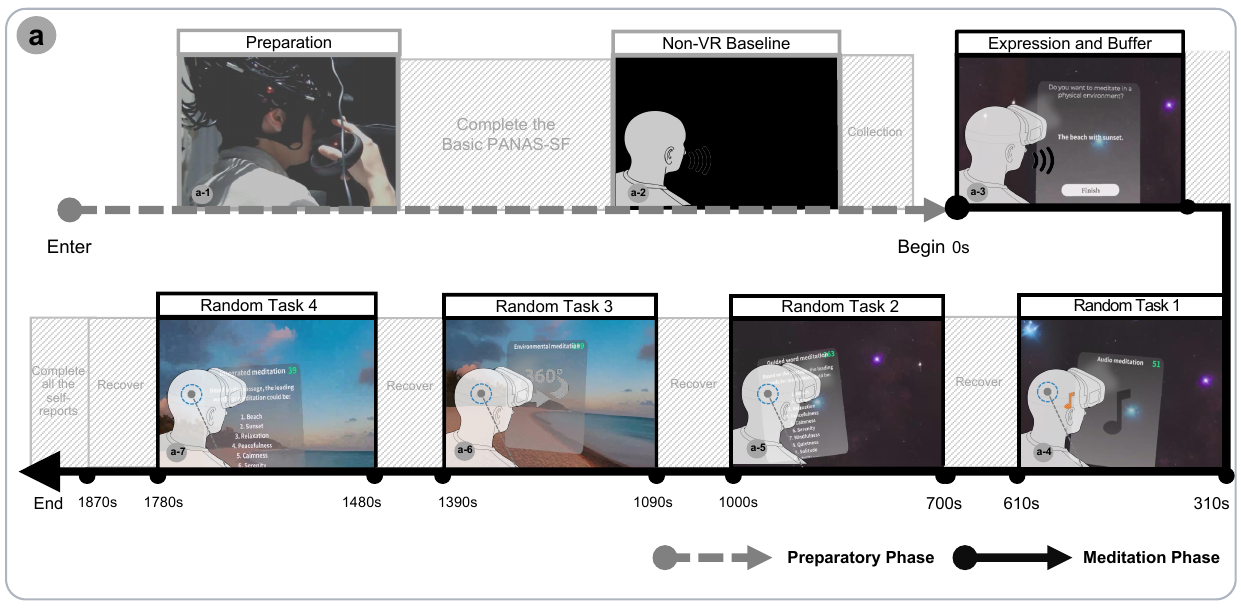}
    \vspace{-25pt}
    \caption{Timeline and task sequence for participant (ID: M11) generative mindfulness tasks in this study. The session begins with a non-VR-based baseline (form a-1 to a-2), followed by (from a-3 to a-6) randomized generative mindfulness tasks interleaved with instrument calibration and rest phases.}
    \label{fig:process}
    \vspace{-10pt}
\end{figure*}

\subsection{Experimental Process}
Our study used a block design within-subject to explore how generative meditation exercises affect cognition, emotion, and brain activity. Key measures included: (1) \textbf{Brain activity}: HbO concentration in predefined ROIs. (2) \textbf{Subjective responses}: Emotional regulation, immersion, and user experience ratings.

\textbf{\textit{Procedure.}} The experiment began with an initial preparation stage, during which the participants were informed of the experimental procedure, followed by simple teaching with VR devices, and then they completed the demographic survey and the short form of the positive and negative affect schedule (PANAS-SF) to establish initial and emotional states. Afterwards, participants were introduced to the experimental procedure and briefly informed about the type of virtual environment they would experience. Meanwhile, the recorded session began with a baseline period (non-VR state) during which participants sat quietly with their eyes closed to stabilize prefrontal hemodynamic signals (see Figure~\ref{sacle}). This served as a baseline for subsequent comparisons in brain analysis~\cite{Bergen2021}.

This experimental paradigm centered on a designed protocol that included four distinct task blocks (Task 1–4), which were interleaved with alternating rest and recovery periods (see Figure~\ref{fig:process}). Each task block introduced a unique form of generative mindfulness meditation, specifically music, verbally guided, integrative, and environmental meditation. Based on this, the experiment was rooted in a "one-factor within-subject design," where a single cohort of participants completed all task conditions, allowing for direct comparison of task-related differences (see Figure~\ref{fig:process}). 

After meditation blocks, during which they completed a series of self-report measures (CHN Version), including the Short Form of the PANAS-SF, the Igroup Presence Questionnaire (IPQ) and the Short Version of the User Experience Questionnaire (UEQ-S), these subjective indicators can observe the emotional changes of the participants after participating in different generative meditations and can also effectively analyze the immersion and user experience quality of the VR frameworks. Moreover, we also conducted semi-structured interviews with participants who completed all tasks, asking questions about their experience and direction for the prototype and system improvement.

\begin{table}[h]
\scriptsize
\centering
\caption{Brain Regions Relevant to Meditation}
\vspace{-10pt}
\begin{tabular}{p{1.1cm}p{3.2cm}p{3.2cm}}
\toprule
\textbf{ROIs} & \textbf{Full Name} & \textbf{Function} \\
\midrule
DLPFC & Dorsolateral Prefrontal Cortex & Cognitive control, attentional regulation and self-monitoring~\cite{Short2007Regional} \\
FPA & Frontal Polar Area & Prospective thinking and self-reflection~\cite{Kral2021} \\
FEF & Frontal Eye Fields & Visual attention and oculomotor control~\cite{Yang2025Comparative} \\
TC & Temporal Cortex & Auditory, semantic and emotional processing~\cite{Newberg2003Cerebral} \\
PreM \& SMC & Premotor and Supplementary Motor Cortex & Somatic awareness, motor planning~\cite{Kwak2020Enhanced}\\
SSC & Somatosensory Cortex & Bodily sensory processing (touch, pressure, proprioception)~\cite{Newberg2001The}\\
\bottomrule

\end{tabular}
\label{tab:brain_regions}
\end{table}

\section{Evaluation Results}

\subsection{Brain Activation (Beta)}
\textbf{\textit{  Data Preprocessing.  }}Brain activity signals were pre-processed, using a standard pipeline: low-quality channel rejection, conversion to HbO/HbR/HbT via the modified Beer-Lambert law, motion correction by sliding-window outlier detection, and zero-phase Butterworth bandpass filtering (0.01--0.20\,Hz). The task segments were then extracted and HbO was used as the primary outcome. We conducted two preregistered primary analyses on \(\Delta\mathrm{HbO}=\beta_{\text{task}}-\beta_{\text{baseline}}\) at the ROI-hemisphere level: (i) an \textit{activation test} for four meditation types—\textbf{environmental meditation}, \textbf{guided verbal meditation}, \textbf{music meditation}, and \textbf{integrative meditation}—assessing whether \(\Delta\mathrm{HbO}>0\) via one-sided paired tests, and (ii) an \textit{across-type comparison} using a one-way repeated-measures ANOVA on \(\Delta\mathrm{HbO}\) followed by paired contrasts. Unless otherwise stated, Bonferroni controlled multiple comparisons within the relevant family, and all reported values \(p\) are corrected values.

To maximize statistical power and align with theory, inference focused on an a priori bilateral ROI set comprising DLPFC, FPA, FEF, TC, PreM\&SMC, and SSC -- these ROIs are all related to the regulatory functions of meditation (see Table~\ref{tab:brain_regions}).

\begin{table}[h]
\setlength{\abovecaptionskip}{2pt}  
\setlength{\belowcaptionskip}{5pt}   
\centering
\begin{threeparttable}
\caption{Activation in ROI of Left and Right (one‑sided paired tests on \(\Delta\)HbO; Bonferroni‑corrected \(p\))}
\label{tab:activation_interestROI_compact}
\scriptsize

\begin{tabular}{l l r r r c}
\toprule
\multicolumn{1}{c}{Tasks} & \multicolumn{1}{c}{ROI--Hem.} & \multicolumn{1}{c}{Mean $\bar{\Delta}$} & \multicolumn{1}{c}{$dz$} & \multicolumn{1}{c}{$F=t^{2}$} & \multicolumn{1}{c}{Sig.} \\
\midrule
\textbf{Environmental meditation} & FEF--R       &  0.028 &  0.490 & 7.840 & \textbf{*} \\
\textbf{Guided verbal meditation} & FEF--R       &  0.025 &  0.460 & 7.010 & $\dagger$  \\
\textbf{Integrative meditation}   & FPA--L       &  0.042 &  0.560 & 9.350 & \textbf{*} \\
\textbf{Integrative meditation}   & FPA--R       &  0.032 &  0.520 & 8.220 & \textbf{*} \\
\textbf{Integrative meditation}   & PreM\&SMC--R &  0.036 &  0.550 & 9.100 & \textbf{*} \\
\bottomrule
\end{tabular}
\begin{tablenotes}[flushleft]
\item \footnotesize \textbf{*} \(p<0.05\);\; $\dagger$ \(0.05<p\le 0.10\). All \(p\) values are Bonferroni‑corrected.
\vspace{-10pt}
\end{tablenotes}
\end{threeparttable}
\end{table}

\textbf{\textit{Activation of Brain Regions.}} Paired one-sided \(t\)-tests on \(\Delta\mathrm{HbO}=\beta_{\text{task}}-\beta_{\text{baseline}}\) revealed robust activation patterns. For \textbf{integrative meditation}, significant increases were observed in \textit{FPA-L} (\(\bar{\Delta}=0.042\), \(dz=0.560\), \(p=0.024\)), \textit{FPA-R} (\(\bar{\Delta}=0.032\), \(dz=0.520\), \(p=0.038\)), and \textit{PreM\&SMC-R} (\(\bar{\Delta}=0.036\), \(dz=0.550\), \(p=0.026\)). \textbf{Environmental meditation} also showed a significant increase in \textit{FEF-R} (\(\bar{\Delta}=0.028\), \(dz=0.490\), \(p=0.048\)). A trend-level increase was found for \textbf{guided verbal meditation} in \textit{FEF-R} (\(\bar{\Delta}=0.025\), \(dz=0.460\), \(p=0.072\)). No other combinations of ROI reached significance or trend thresholds (see Table~\ref{tab:activation_interestROI_compact}). These findings indicate that multimodal content engages the frontopolar and premotor-sensorimotor regions, while environmental signals can up-regulate activity in the frontal eye fields. They also highlights that personalized and generative meditation requires greater attention to the process of integration.

\begin{figure}[h]
    \centering
    \includegraphics[width=1\linewidth]{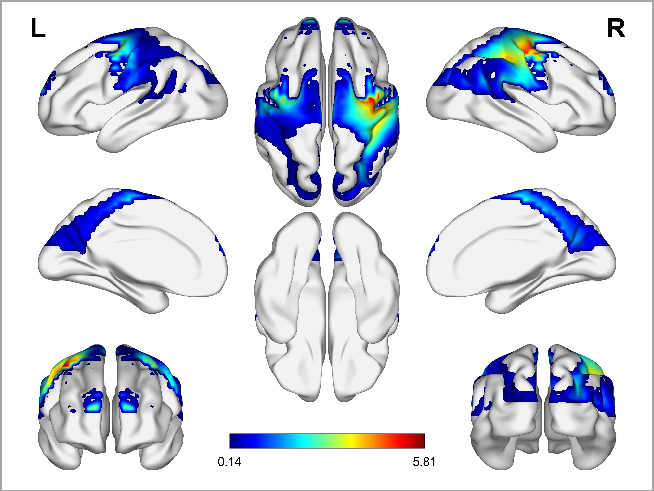}
    \vspace{-10pt}
    \caption{Surface projection of the main effect of  four tasks in L/R mark hemispheres. The one-way repeated measures ANOVA was performed on the four meditation tasks. }
    \vspace{-10pt}
    \label{fig:naotu}
\end{figure}

\textbf{\textit{Generative Meditations Comparison.}} In this analysis, a one-way repeated measure ANOVA was performed on the four meditation tasks for each combination of ROI-hemisphere (see Table~\ref{tab:anova_main} and Figure~\ref{fig:naotu}). Significant main effects emerged in \textit{PreM\&SMC-R} (\(F(3,87)=5.812,\ p=0.011,\ \text{partial }\eta^2=0.167\)) and a trend in \textit{PreM\&SMC-L} (\(F(3,87)=4.172,\ p=0.082,\ \text{partial }\eta^2=0.126\)). Bonferroni-adjusted post hoc tests showed that \textbf{integrative meditation} elicited higher activation than \textbf{music meditation} (\(\Delta=0.035,\ dz=0.700,\ p=0.004\)) and \textbf{guided verbal meditation} (\(\Delta=0.035,\ dz=0.540,\ p=0.038\)) in \textit{PreM\&SMC-R}. A trend was observed in \textit{PreM\&SMC-L} for the music vs.\ integrative contrast (\(\Delta=0.043,\ dz=0.480,\ p=0.080\)). None of the other comparisons met the significance or trend thresholds (\(p>0.10\)), reinforcing the selective involvement of the premotor and supplementary motor cortex in responses to integrative content (see Table~\ref{tab:posthoc}).

\begin{table}[h]
\caption{Repeated-Measures ANOVA on $\Delta$HbO Across Meditation Types (Bonferroni-corrected \(p\))}
\label{tab:anova_main}
\vspace{-10pt}
\centering
\scriptsize 
\begin{minipage}{\linewidth}
\begin{threeparttable}
\setlength\tabcolsep{8pt} 
\resizebox{\linewidth}{!}{%
\begin{tabular}{lcccc}
\toprule
\multicolumn{1}{c}{ROI--Hem.} & \multicolumn{1}{c}{$F$} & \multicolumn{1}{c}{$p$} & \multicolumn{1}{c}{$\eta^2_{p}$} & \multicolumn{1}{c}{Sig.} \\
\midrule
DLPFC--L       & 1.120  & 0.345 & 0.038 &        \\
DLPFC--R       & 0.950  & 0.420 & 0.032 &        \\
FPA--L         & 1.580  & 0.202 & 0.052 &        \\
FPA--R         & 1.460  & 0.233 & 0.048 &        \\
TC--L          & 0.880  & 0.456 & 0.030 &        \\
TC--R          & 1.040  & 0.377 & 0.035 &        \\
PreM\&SMC--L   & 4.172  & 0.082 & 0.126 & $\dagger$ \\
PreM\&SMC--R   & 5.812  & 0.011 & 0.167 & \textbf{*} \\
SSC--L         & 0.710  & 0.550 & 0.024 &        \\
SSC--R         & 0.760  & 0.517 & 0.026 &        \\
\bottomrule
\end{tabular}%
}
\begin{tablenotes}[flushleft]
\scriptsize
\item \textbf{*} \(p\le 0.05\);\; $\dagger$ \(0.05<p\le 0.10\).
\vspace{-10pt}
\end{tablenotes}
\end{threeparttable}
\end{minipage}
\end{table}

After correction for multiple comparisons, integrative meditation elicited significantly higher ΔHbO in the bilateral frontopolar area (FPA) and the right premotor and supplementary motor cortex (PreM \& SMC), with a trend observed in the left PreM \& SMC. According to the functional roles summarized in Table~\ref{tab:brain_regions}, activation in FPA - associated with prospective thinking and self-reflection - suggests that integrative meditation effectively engages higher-order self-regulation processes. Concurrently, the PreM \& SMC, linked to somatic awareness and motor planning, exhibited enhanced activation, indicating increased bodily engagement and internal monitoring. These results suggest that multimodal integration supports both cognitive and embodied components of meditation, thus enhancing self-regulation through coordinated activation across reflective and sensorimotor systems (see Figure~\ref{fig:naotu}). This neurophysiological profile provides empirical support for RQ1, highlighting the distinct advantage of generative multisensory meditation in activating relevant brain circuits.

\begin{table}[t]
\setlength{\abovecaptionskip}{2pt}
\setlength{\belowcaptionskip}{4pt}
\centering
\begin{threeparttable}
\caption{Post Hoc Pairwise Comparisons in Meditation Tasks (Bonferroni‑corrected \(p\))}
\label{tab:posthoc}
\footnotesize
\setlength\tabcolsep{5pt}
\begin{tabular}{l l r r c}
\toprule
\multicolumn{1}{c}{ROI--Hem.} & \multicolumn{1}{c}{Tasks Comparison} &
\multicolumn{1}{c}{Mean $\Delta$} & \multicolumn{1}{c}{$dz$} &
\multicolumn{1}{c}{\(p\)} \\
\midrule
FPA--L            & music vs.\ integrative           &  -0.027 &  -0.520 & 0.048 \textbf{*} \\
FPA--R            & guided verbal vs.\ integrative   &  -0.021 &  -0.540 & 0.037 \textbf{*} \\
PreM\&SMC--L      & music vs.\ integrative           &  -0.043 &  -0.480 & 0.080 $\dagger$ \\
PreM\&SMC--R      & music vs.\ integrative           &  -0.035 &  -0.700 & 0.004 \textbf{*} \\
PreM\&SMC--R      & guided verbal vs.\ integrative   &  -0.035 &  -0.540 & 0.038 \textbf{*} \\
\bottomrule
\end{tabular}
\begin{tablenotes}[flushleft]
\item \footnotesize  \textbf{*} \(p<0.05\);\; $\dagger$ \(0.05<p\le 0.10\).

\end{tablenotes}
\end{threeparttable}
\vspace{-10pt}
\end{table}


\subsection{Emotional Changes}

Participants completed a non-VR baseline and four VR-based generative tasks. Condition-wise changes were analyzed with paired Wilcoxon signed-rank tests comparing each task to baseline, and bonferroni correction was applied within item families (reported \(p\) values are corrected).

\begin{figure}[h]
    \centering
    \includegraphics[width=1\linewidth]{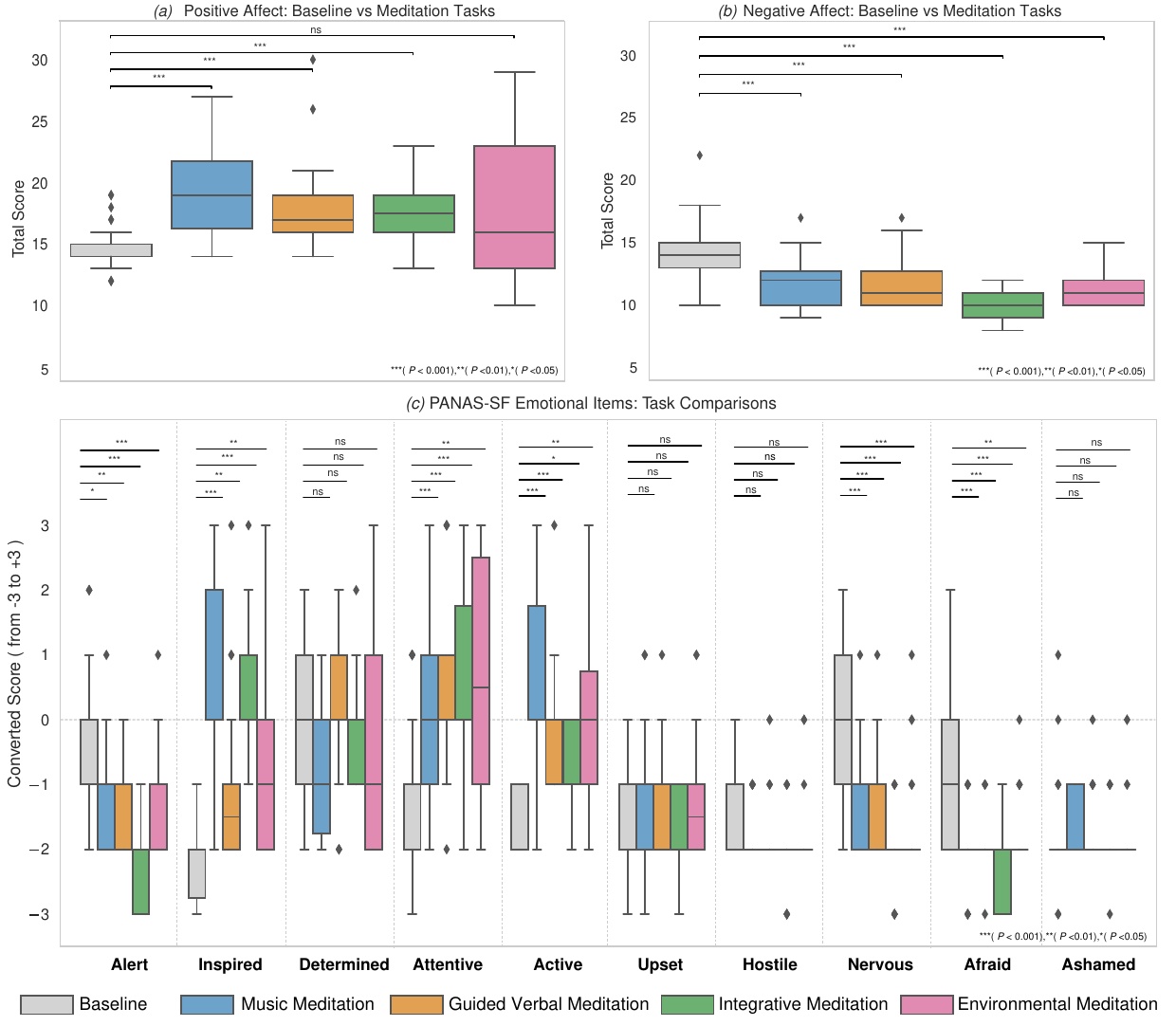}
    \vspace{-10pt}
    \caption{PANAS‑SF results. Panels (a) and (b) show PA and NA changes from baseline to each meditation task; panel (c) shows item‑level comparisons.}
    \vspace{-20pt}
    \label{PANAS-FS}
\end{figure}

\textbf{\textit{Changes in Positive and Negative Emotions.}} Total PA increased significantly after \textbf{music meditation} (\(p<0.001\)), \textbf{guided verbal meditation} (\(p<0.001\)), and \textbf{integrative meditation} (\(p<0.001\)), with \textbf{environmental meditation} showing a trend (\(p=0.061\)). In contrast, all four tasks robustly reduced NA: \textbf{music meditation} (\(p<0.001\)), \textbf{guided verbal meditation} (\(p=0.001\)), \textbf{integrative meditation} (\(p<0.001\)), and \textbf{environmental meditation} (\(p=0.001\)). Specifically, \textbf{integrative meditation} produced the strongest combined profile: substantially improving PA while, together with \textbf{environmental meditation}, yielding the largest NA decreases—supporting broad benefits to affect regulation (see Figure~\ref{PANAS-FS}).

\textbf{\textit{Changes in Specific Emotional Items.}} Positive items showed widespread increases: \textit{Alert}, \textit{Inspired}, and \textit{Attentive} rose across all tasks (\(p<0.001\) for most comparisons). \textit{Determined} increased for \textbf{music meditation} (\(p=0.013\)), \textbf{guided verbal meditation} (\(p=0.001\)) and \textbf{integrative meditation} (\(p=0.001\)), with a marginal increase for \textbf{environmental meditation} (\(p=0.034\)). \textit{Active} increased for all tasks except \textbf{environmental meditation} (\(p>0.050\)). Negative items decreased reliably: \textit{Afraid} declined across all tasks (\(p\le 0.002\)), while \textit{Nervous} and \textit{Upset} decreased during \textbf{music}, \textbf{guided verbal}, and \textbf{integrative meditation} (\(p\le 0.001\)). \textit{Hostile} decreased for \textbf{guided verbal meditation} (\(p=0.001\)), \textbf{integrative meditation} (\(p=0.001\)), and \textbf{environmental meditation} (\(p=0.004\)). \textit{Ashamed} remained unchanged in all tasks (\(p>0.050\)), consistent with a floor effect. Overall, VR-based generative meditation reduced negative emotions (especially fear, nervousness, and upset) and increased positive emotions (notably inspiration, alertness, and attention), with the most pronounced PA gains during \textbf{integrative meditation} and other content-rich modes (RQ2).

\subsection{Immersion and Experience in VR}
\begin{figure}[h]
    \centering
    \includegraphics[width=1\linewidth]{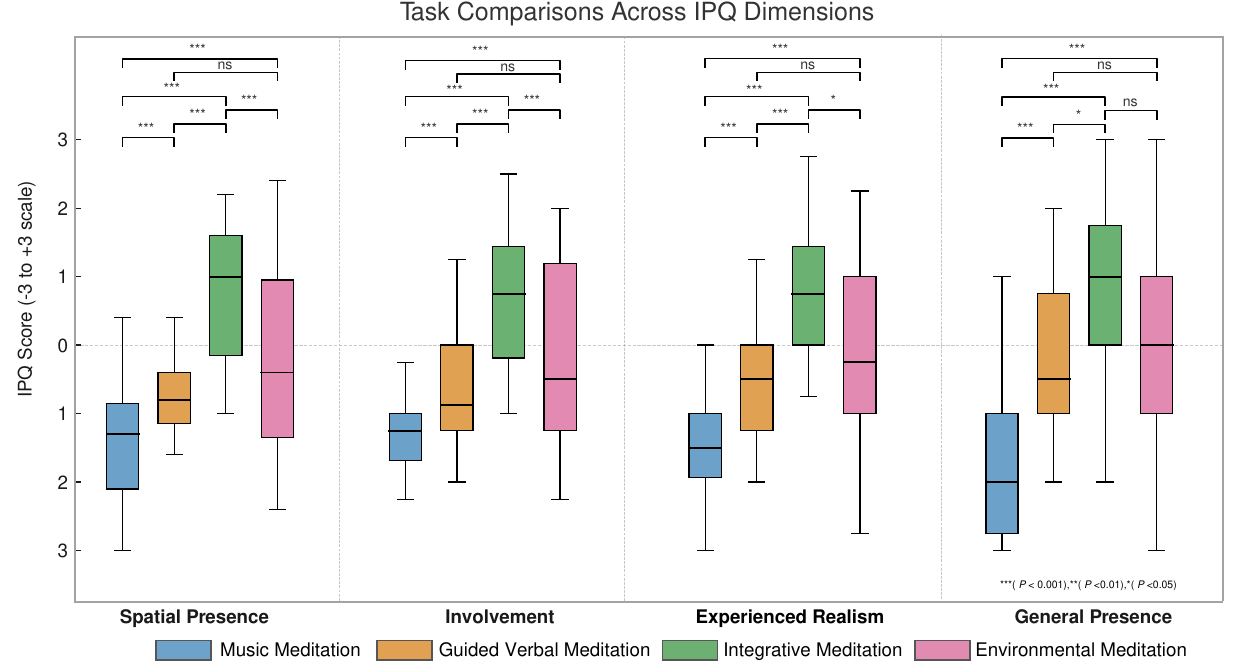}
    \vspace{-10pt}
    \caption{IPQ results across four dimensions: Spatial, Involvement, Experienced Realism, and General Presence.}
        \vspace{-10pt}
    \label{IPQ}
\end{figure}

\quad \textbf{\textit{Igroup Presence Questionnaire (IPQ).}} ANOVAs (Repeated measures) with Greenhouse-Geisser correction showed a significant main effect of Task on all IPQ dimensions (all \(p<0.001\)): Spatial Presence, \(F(3,87)=39.190\); Involvement, \(F(3,87)=30.580\); Experienced Realism, \(F(3,87)=30.680\); and General Presence, \(F(3,87)=19.440\) (see Figure~\ref{IPQ}). Descriptively, \textbf{integrative meditation} yielded the highest scores across subscales (Spatial Presence: \(M=0.800\); Involvement: \(M=0.690\); Experienced Realism: \(M=0.670\); General Presence: \(M=0.600\)), while \textbf{music meditation} produced the lowest. Bonferroni-corrected pairwise comparisons indicated that \textbf{integrative meditation} exceeded \textbf{music meditation} and \textbf{guided verbal meditation} on all subscales (all \(p<0.001\)), and exceeded \textbf{environmental meditation} on Spatial Presence (\(p<0.001\)), Involvement (\(p<0.001\)), and Experienced Realism (\(p=0.015\)). In contrast, \textbf{music meditation} scored significantly lower than all other conditions on every dimension (all \(p\le 0.001\)). \textbf{Guided verbal meditation} and \textbf{environmental meditation} showed comparable presence levels (Spatial Presence: \(p=0.420\); Involvement: \(p=0.420\); Experienced Realism: \(p=0.420\)). For General Presence, \textbf{integrative meditation} and \textbf{environmental meditation} did not differ (\(p=0.330\)), nor did \textbf{guided verbal meditation} and \textbf{environmental meditation} (\(p=1.000\)). Together, multimodal \textbf{integrative meditation} robustly enhances spatial presence, participation, and perceived realism, while unimodal \textbf{ music meditation} yields the weakest experience of presence (RQ3).

\textbf{\textit{User Experience.}} This evaluation revealed systematic differences across tasks for Pragmatic Quality (PQ: support, efficiency, clarity) and Hedonic Quality (HQ: innovation, appeal). All scales demonstrated excellent reliability: PQ (\(\alpha=0.870\)), HQ (\(\alpha=0.860\)), and overall user experience (\(\alpha=0.910\)). Greenhouse-Geisser corrected ANOVAs showed strong main effects with large effect sizes: PQ, \(F(2.44,70.66)=42.420\), \(p<0.001\), \(\eta^2_p=0.590\); HQ, \linebreak \(F(2.64,76.58)=44.030\), \(p<0.001\), \(\eta^2_p=0.600\); overall user experience: \(F(2.61,75.65)=58.210\), \(p<0.001\), \(\eta^2_p=0.670\). Descriptively, \textbf{integrative meditation} scored highest across measures (PQ: \(M=1.070\); HQ: \(M=1.080\); overall user experience: \(M=1.070\)), while \textbf{music meditation} (PQ: \(M=-0.450\); HQ: \(M=-0.940\); overall user experience: \(M=-0.700\)) and \textbf{environmental meditation} (PQ: \(M=-0.930\); HQ: \(M=-0.990\); overall: \(M=-0.960\)) were lowest. \textbf{Guided verbal meditation} showed intermediate values (PQ: \(M=1.020\); HQ: \(M=0.120\); overall: \(M=0.570\)).

\begin{figure}[h]
    \centering
    \includegraphics[width=1\linewidth]{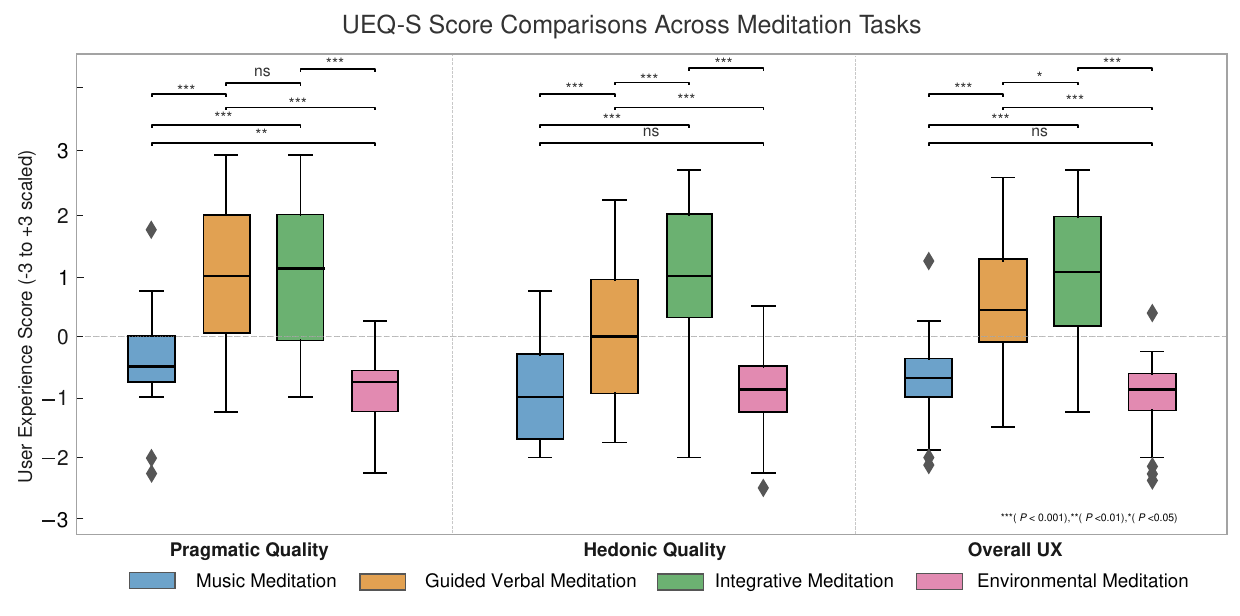}
    \vspace{-10pt}
    \caption{UEQ‑S results on the Pragmatic Quality and Hedonic Quality in the comparison of meditation tasks. }
    \label{UEQ}
    \vspace{-10pt}
\end{figure}

Pairwise comparisons (Holm-adjusted) confirmed the superiority of \textbf{integrative meditation}: it significantly outperformed all other tasks in HQ and overall user experience (all \(p<0.001\)), and surpassed both \textbf{music meditation} and \textbf{environmental meditation} in PQ (\(p<0.001\)). Critically, it exceeded \textbf{guided verbal meditation} in overall user experience (\(p=0.046\)). \textbf{Guided verbal meditation} consistently outperformed \textbf{music meditation} and \textbf{environmental meditation} in PQ and overall user experience (all \(p<0.001\)). No differences emerged between \textbf{music meditation} and \textbf{environmental meditation} for HQ (\(p=0.800\)) or overall user experience (\(p=0.078\)). Strong correlations supported the composite measure, with overall user experience closely linked to both PQ and HQ (\(r=0.930\), \(p<0.001\) each; see Figure~\ref{UEQ}). Collectively, \textbf{integrative meditation} emerges as the optimal condition, while unimodal tasks underperform.

\textbf{\textit{Insights from Semi-Structured Interviews.}} The interviews revealed nuanced perspectives on AI-driven generative mindfulness systems compared to traditional approaches. Thematic analysis highlighted three opportunities for improvement.

\begin{itemize}
    \item Need for real-time feedback: The synchronization of meditation with physiological signals resonated with several participants. As one participant noted, \textit{"If I could see my real-time heart rate variability aligning with the AI narrative changes, I would trust the system more"} (M13).
    \item Clinical-grade personalization: The participants emphasized adaptation to individual neurocognitive needs. One neurodiverse user shared, \textit{"My brain needs shorter, reward-based sessions that rigid timelines cannot provide, but AI could learn this"} (M09).
\end{itemize}


Synthesizing IPQ, UEQ-S, and interview data indicates that multisensory integration fundamentally shapes both the presence and user experience in VR meditation (RQ3). In particular, \textbf{integrative meditation}—synergizing visual, auditory, and verbal stimuli - consistently outperformed unimodal approaches.

\section{Discussion and Conclusion}
\textbf{\textit{Psychological intervention.}} \textbf{Integrative generative meditation} robustly activated bilateral frontopolar cortex (FPA) and right premotor/supplementary motor cortex (PreM \& SMC), while \textbf{environmental meditation} engaged the right frontal eye field (FEF). Functionally, FPA supports prospective thinking and self-regulation, whereas PreM \& SMC underlies somatic awareness and motor preparation. This co-activation suggests that integrative meditation fosters a cognitively embodied state, combining forward-oriented internal simulation with heightened bodily monitoring—potentially explaining users' reported sense of presence and agency. Conversely, FEF activation during environmental meditation indicates enhanced visual attention in spatially rich VR scenes. These findings indicate that dynamic multisensory content scaffolds distinct regulatory mechanisms: \textbf{integrative meditation} harmonizes top-down reflective processes (FPA) with bottom-up sensorimotor grounding (PreM \& SMC), offering a neurophysiological pathway for short-term psychological self-regulation~\cite{Singh2022}.

\textbf{\textit{Emotional Regulation.}}
Across generative conditions, the negative affect decreased strongly, while the positive affect increased after \textbf{music meditation}, \textbf{guided verbal meditation}, and \textbf{ integrated meditation}. Thus, the strongest combined affective profile was achieved by \textbf{ integrated meditation}, which both enhanced PA and—together with \textbf{environmental meditation}—produced the NA reductions. These results are consistent with the idea that various meditation formats act through overlapping emotion regulation mechanisms~\cite{Ameta2023,Han2023}. At the same time, experiential measures qualify the affective findings: \textbf{ music meditation} increased PA, but produced the weakest presence and user experience, suggesting that music alone may not provide sufficient spatial/narrative scaffolding. \textbf{Guided verbal meditation} showed intermediate user experience and presence, consistent with its role in supporting goal‑directed regulation and attentional focus, while \textbf{environmental meditation}—despite its strong negative emotions down-regulation. These patterns argue for coupling affective targets with interactional structure when designing VR‑based interventions.

\textbf{\textit{Multimodal Integration.}} The results highlight the power of multimodal integration to shape immersive and satisfying VR experiences. Integrated meditation, which combines visual elements, ambient sound, and guided verbal cues, created a sensory environment that users perceived as more coherent and believable. Relatively, music-guided meditation performed worse in all dimensions. As one participant (F04) said, ``\textit{Listening to music in VR feels `clumsy'.}” -- it highlights how disjointed the experience can be when audio lacks spatial context or interactive cues. Without perceptual anchoring or narrative direction, music alone often does not support immersion, leaving users feeling detached rather than engaged.

\textbf{\textit{Summary, Limitations, and Future Work.}}  MindfulVerse demonstrates how AI can generically and dynamically create adaptive multisensory meditation experiences by synthesizing real-time environments, audio generation, and personalized guidance.  These outcomes collectively validate the technical feasibility and experiential efficacy of AI-generated meditation, highlighting design opportunities to tailor modality-specific implementations based on user needs and context.

Several limitations tempered these conclusions. First, large scene generation introduced network latency that undermined immersion. Second, customization was constrained by limited control over the visual style and the parameters of the model. Third, the interaction remained largely static because user state signals were not yet incorporated in a closed-loop manner. Finally, the evaluation was short-term, leaving the longitudinal efficacy of generative meditation relative to conventional VR mindfulness untested. 

To address these issues, future work will pursue three interconnected directions. Rigorous longitudinal comparisons of traditional VR meditation and generative meditation will evaluate their differential impacts on engagement and affect regulation, building on prior studies~\cite{lee2024impact,gromala2015virtual,kober2024controlling}. Architecturally, we will shift from remote to local or hybrid rendering models while expanding user control over generative style and parameters to achieve personalized aesthetics. Most critically, real-time adaptation mechanisms will be developed to fuse behavioral, affective, and physiological signals—including brain activity~\cite{Pan2023}—enabling neuroadaptive content tailored to dynamic states such as stress, focus, and relaxation~\cite{Tran2024}. Together, these advances will ground next-generation VR wellness systems in empirical validation while achieving scalable personalization.

\begin{acks}
This work was supported in part by the China National Key Research and Development Program under Grant 2024YFC3307602, the Guangzhou Municipal Nansha District Science and Technology Bureau under Contract No.2022ZD01.  
\end{acks}

\balance
\bibliographystyle{ACM-Reference-Format}
\bibliography{samples/software}

\end{document}